\documentclass[prb,showpacs,floatfix,twocolumn,superscriptaddress]{revtex4}
\usepackage{graphicx}
\usepackage{dcolumn}
\begin{document}


\title{Evidence for the band broadening across the ferromagnetic transition in Cr$_{1/3}$NbSe$_2$}
\author{W. Z. Hu}
\author{G. T. Wang}
\affiliation{Beijing National Laboratory for Condensed Matter
Physics, Institute of Physics, Chinese Academy of Sciences,
Beijing 100080, People's Republic of China}
\author{Rongwei Hu}
\affiliation{Condensed Matter Physics and Materials Science
Department, Brookhaven National Laboratory, Upton, New York 11973,
USA} \affiliation{Physics Department, Brown University, Providence
RI 02912, USA}
\author{C. Petrovic}
\affiliation{Condensed Matter Physics and Materials Science
Department, Brookhaven National Laboratory, Upton, New York 11973,
USA}
\author{E. Morosan}
\author{R. J. Cava}
\affiliation{Department of Chemistry, Princeton University,
Princeton, New Jersey 08540, USA}
\author{Z. Fang}
\affiliation{Beijing National Laboratory for Condensed Matter
Physics, Institute of Physics, Chinese Academy of Sciences,
Beijing 100080, People's Republic of China}
\author{N. L. Wang}
\email{nlwang@aphy.iphy.ac.cn}%
\affiliation{Beijing National Laboratory for Condensed Matter
Physics, Institute of Physics, Chinese Academy of Sciences,
Beijing 100080, People's Republic of China}
%


\begin{abstract}
The electronic structure of Cr$_{1/3}$NbSe$_2$ is studied via
optical spectroscopy. We observe two low-energy interband
transitions in the paramagnetic phase, which split into four peaks
as the compound enters the ferromagnetic state. The band structure
calculation indicates the four peaks are interband transitions to
the spin up Cr e$_g$ states. We show that the peak splitting below
the Curie temperature is \emph{not} due to the exchange splitting
of spin up and down bands, but directly reflects a band broadening
effect in Cr-derived states upon the spontaneous ferromagnetic
ordering.
\end{abstract}


\maketitle

\section{Introduction}
Layered transition metal dichalcogenides TX$_2$ (T=transition
metal, X=chalcogen) are among the most studied two-dimensional
electronic systems. The charge-density-wave (CDW) instability and
its coexistence/competition with superconductivity are central
characteristics of this family.\cite{Morosan, NaxTaS2} On the
other hand, due to the weak van der Waals interaction between
X-T-X sandwich layers, a large variety of atoms and molecules can
be intercalated into the interlayer vacant sites, and dramatically
change the physical properties.\cite{TMDC, Friend MxTX2, Parkin,
Fe1/4TaS2}

Among various intercalates, the 3\emph{d}-transition metal
intercalated 2$H$-type M$_x$TX$_2$ are particularly interesting
(M=V, Cr, Mn, Fe, Co, Ni; 2$H$ means \emph{two} X-T-X sandwiches
in one unit cell, and the compound has a \emph{hexagonal}
symmetry). At 1/4 or 1/3 doping, the intercalated 3\emph{d} ions
form an ordered two dimensional magnetic array, and show either
ferromagnetic (FM) or antiferromagnetic (AFM) order at low
\emph{T}, depending on the intercalated species and the host
compound.\cite{Friend MxTX2, Parkin, Fe1/4TaS2} Transport
measurements commonly indicate metallic conductivity in the
paramagnetic (PM) state and a more rapid decrease in resistivity
in the spin ordered state. The easy axis of the magnetization can
be either parallel or perpendicular to the hexagonal layers for
different FM intercalates.\cite{Friend MxTX2,Parkin} The types of
magnetic ordering encountered provide an interesting venue for
exploring spin-related effects, for example, the anomalous Hall
effect\cite{Fe1/4TaS2 Hall} or the magneto-optical effect, topics
of renewed and increased interest due to the recent development of
intrinsic mechanism arising from the Berry phase of the Bloch
state.\cite{Yao}

\begin{figure}[b]
\includegraphics[width=2.5in]{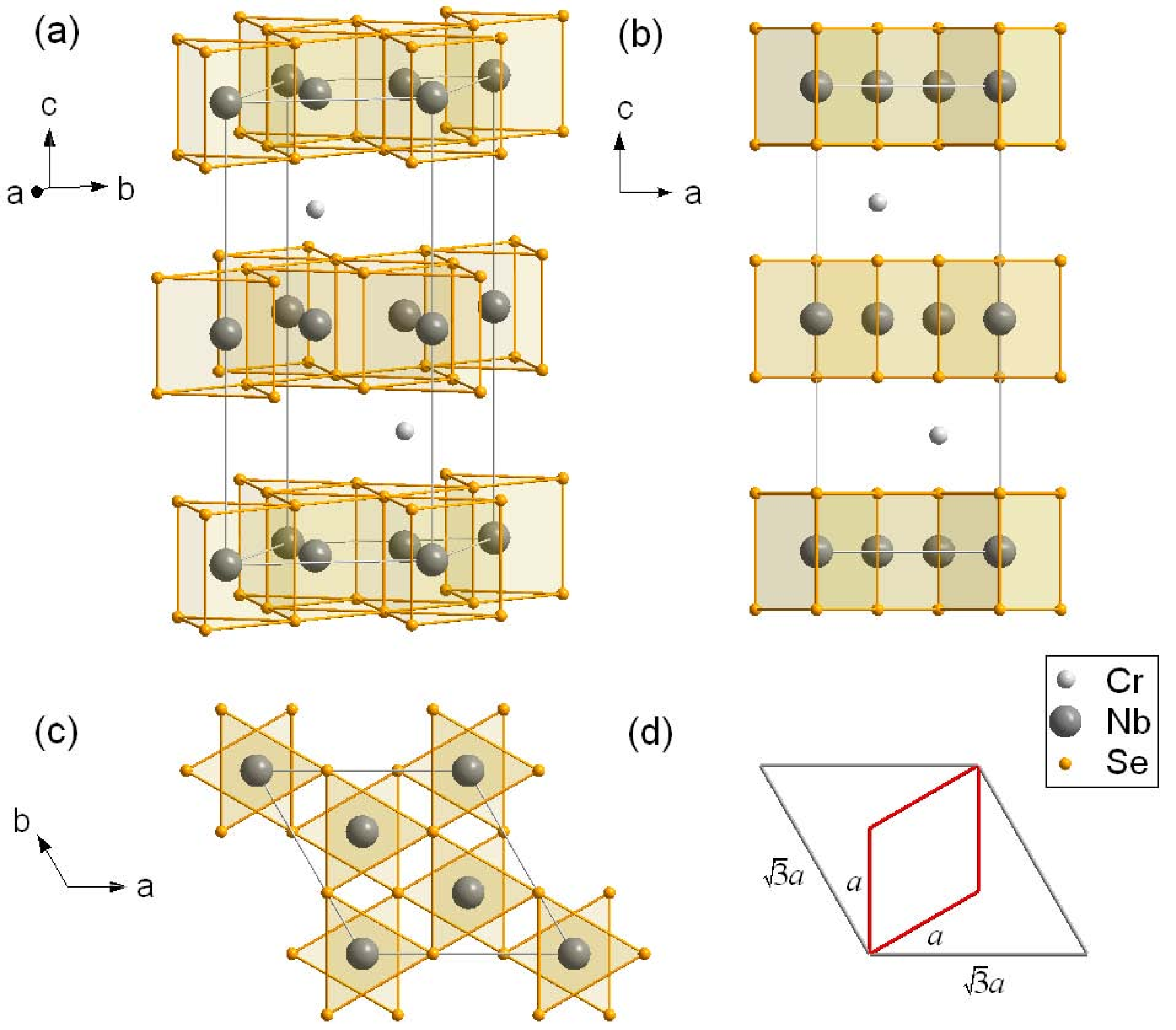}%
\vspace*{-0.20cm}%
\caption{\label{fig:structure} (Color online) The unit cell of
2$H$-Cr$_{1/3}$NbSe$_2$: (a) a 3D view; (b) a (010) projection: Cr
atoms occupy one third of the octahedral holes in the van der
Waals gap; (c) a (001) projection; (d) a comparison of the
NbSe$_2$ unit cell (red) and the Cr$_{1/3}$NbSe$_2$
$\sqrt{3}\times\sqrt{3}$ superlattice (grey).}
\end{figure}

The \emph{T}-dependent band structure, especially its modification
across the FM transition, provides essential information in
understanding ferromagnetism. In the itinerant Stoner
model,\cite{Wohlfarth} the energy gain in the FM state arises from
the exchange interaction, and the collapse of exchange splitting
leads to a vanishing magnetic moment above the Curie \emph{T}.
While in the localized model,\cite{Anderson} ferromagnetism
originates from the long-range ordered local moments, which are
orientation disordered in the PM state. However, practical
ferromagnets usually belong to the intermediate regime between the
above two extremes, requiring a unified picture which covers the
whole range from itinerant to localized
ferromagnetism.\cite{Moriya}

In this work, we present an optical spectroscopy study and
first-principles calculations on Cr$_{1/3}$NbSe$_2$ to elucidate
the electronic structure change across the FM transition. In
comparison with pure 2$H$-NbSe$_2$, two new interband transitions
in the mid- and near-infrared region emerge in Cr$_{1/3}$NbSe$_2$,
which split into four in the FM state. Band structure calculation
indicates a strong hybridization between Cr 3\emph{d} and Nb
4\emph{d} bands near E$_F$, and the spin up Cr e$_g$ shows a
double-peak character in the density of states (DOS) map. Then the
four optical peaks in the FM state are all interband transitions
to Cr e$_g$($\uparrow$), and the band broadening in this
Cr-derived state is the cause for the peak splitting in the
optical response.

\section{Details for Experiment and Band Structure Calculations}
Both 2$H$-NbSe$_2$ and 2$H$-type Cr$_{1/3}$NbSe$_2$ single
crystals were grown by the vapor transport method. The structure
of Cr$_{1/3}$NbSe$_2$ is shown in Fig. 1. The \emph{T}-dependent
resistivity was obtained by the four contact technique in a
Quantum Design PPMS. The near-normal incident reflectance spectra
were measured by a Bruker IFS 66v/s spectrometer in the frequency
range from 40 to 25000 cm$^{-1}$. An \textit{in situ} gold and
aluminium overcoating technique was used to get the reflectivity
R($\omega$). The real part of conductivity $\sigma_1(\omega)$ is
obtained by the Kramers-Kronig transformation of R($\omega$).

First-principles calculations based on the local density
approximation (LDA) and local spin density approximation (LSDA)
were performed to get the electronic structure of NbSe$_2$ and
Cr$_{1/3}$NbSe$_2$, respectively. The calculations were done with
our STATE code.\cite{code} We adopt the experimental structure
parameters\cite{parameter}, and an initial Se position
z=0.125\emph{c} for both compounds, then relax the Se ions to
their lowest energy positions (z$_{Se}$=0.132\emph{c} for
2$H$-NbSe$_2$; z$_{Se}$=0.131\emph{c} for Cr$_{1/3}$NbSe$_2$). In
previous calculation, z$_{Se}$ for 2$H$-NbSe$_2$ is around
0.118\emph{c}\cite{NbSe2 band}. The discrepancy arises from
different choices for the atomic coordination. Here we set the Nb
position as (0,0,0), not (0,0,1/4\emph{c}).\cite{NbSe2 band} Note
z$_{Se}$=0.132\emph{c} in Nb(0,0,0) is in fact equivalent to
z$_{Se}$=0.118\emph{c} in Nb(0,0,1/4\emph{c}) coordination.

\section{Experimental Results: Transport and Optical Properties}

As shown in Fig. 2, a metallic behavior is found for
Cr$_{1/3}$NbSe$_2$ from 300 to 10 K. The resistivity drops more
rapidly with decreasing \emph{T} below 115 K, consistent with the
onset temperature of FM ordering.\cite{Cr1/3NbSe2, Parkin} Similar
transport behavior is observed in other 3\emph{d} transition metal
intercalates.\cite{Friend MxTX2, Fe1/4TaS2, Parkin} The inset of
Fig. 2 focuses on the CDW and superconducting transitions for
2$H$-NbSe$_2$.

\begin{figure}[t]
\includegraphics[width=2.3in]{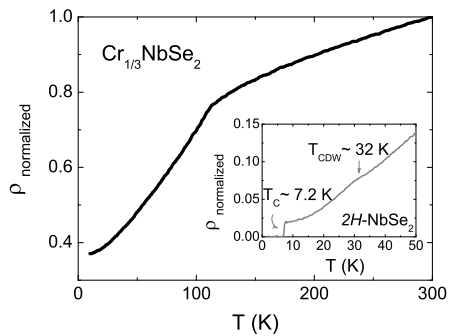}%
\vspace*{-0.20cm}%
\caption{\label{fig:resistivity} Normalized \emph{T}-dependent
resistivity for Cr$_{1/3}$NbSe$_2$. Inset: an expanded plot for
the CDW ($\sim$32 K) and superconducting transitions ($\sim$7.2 K)
in 2$H$-NbSe$_2$.}
\end{figure}

The R($\omega$) for 2$H$-NbSe$_2$ (Fig. 3 (a)) are the same as
found in a previous study,\cite{FT-IR on NbSe2} except for a new
feature in the mid-infrared region: R($\omega$) at 10 K is
suppressed slightly below that at 50 K from 600 to 5000 cm$^{-1}$.
This is intrinsic, since the same result was repeatedly obtained,
with each study carried out on a fresh, newly cleaved surface. The
midinfrared suppression implies the formation of a partial energy
gap on the Fermi surface in the CDW state, which closely resembles
that of 2$H$-TaS$_2$\cite{TaS2 optic}, another CDW-bearing member
with a higher \emph{T$_{CDW}$} (75 K) and thus a more apparent gap
character.

R($\omega$) is greatly modified after Cr intercalation (Fig. 3
(b)). The well-defined plasma edge for 2$H$-NbSe$_2$ decays into a
overdamped one in Cr$_{1/3}$NbSe$_2$. Such an overdamped shape
might arise from impurity scattering due to Cr disorder, but not
the scattering by Cr moments, since the shape of the plasma edge
is almost unchanged from FM to PM states. The most interesting
features are two additional peaks around 4300 and 8500 cm$^{-1}$
at 300 K, and their splitting below 115 K. No similar feature is
found in the host compound.

\begin{figure}[t]
\includegraphics[width=2.5in]{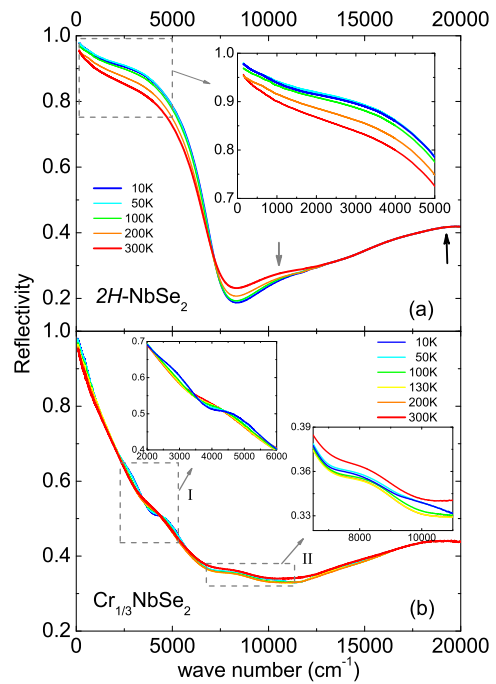}%
\vspace*{-0.20cm}%
\caption{\label{fig:R cmp}(Color online) Reflectivity of (a)
2$H$-NbSe$_2$ and (b) Cr$_{1/3}$NbSe$_2$ below 20000 cm$^{-1}$ at
various temperatures. The inset figures amplify the low frequency
characters.}
\end{figure}

The low energy interband transitions in Cr$_{1/3}$NbSe$_2$ can be
more clearly resolved in the real part of the conductivity
$\sigma$$_1$($\omega$), as illustrated in Fig. 4 (a). Beside the
Drude component, two peaks around 4300 and 8500 cm$^{-1}$ at 300 K
split into four peaks near 2600, 4700, 8400 and 10500 cm$^{-1}$ in
the FM state. In order to extract these features from the
background, we fit $\sigma$$_1$($\omega$) by the Drude-Lorentz
model.\cite{DL model} In the inset of Fig. 4 (a), the fitting
curves of 10 and 300 K match the original data well. Subtracting
the Drude components and the high energy interband transitions
from $\sigma$$_1$($\omega$), we then get a clear picture of the
absorption peaks and their splitting in Fig. 4 (b): the peak
$\delta_1$ (4250 cm$^{-1}$) at 300 K splits into $\alpha_1$ (2610
cm$^{-1}$) and $\beta_1$ (4690 cm$^{-1}$) at 10 K; the peak
$\delta_2$ (8840 cm$^{-1}$) splits into $\alpha_2$ (8400
cm$^{-1}$) and $\beta_2$ (10520 cm$^{-1}$). The splitting energies
$\beta_1-\alpha_1$ and $\beta_2-\alpha_2$ are almost the same,
hence those peaks should have a similar origin. The apparent
spectral difference for NbSe$_2$ and Cr$_{1/3}$NbSe$_2$ indicates
Cr intercalation is not a simple charge transfer process. Detailed
information about the band structure is required.

We note that the spectral difference (i.e. the splitting of the
interband transitions) at 100 K and 130 K in the near-infrared
region is obscure. This is understandable as those features in
R($\omega$) are close to the reflectivity minimum and the overall
peak strengths are rather weak. In addition, both T=100 K and 130K
are very close to the FM transition temperature (115K), therefore
fluctuation effect is also prone to influence the spectra near the
critical temperature. On the other hand, in our overcoating
technique for reflectance measurement, the sample is not in
exchange gas but in vacuum (better than 3$\times10^{-7}$ torr),
the measurement becomes relatively surface sensitive in such high
energy, a more detailed inspection of temperature variations in
$\sigma_1(\omega)$ for the FM phase, which requires to keep the
sample at low temperature for a long time, becomes impractical.
However, the splitting indeed turns more apparent as the
temperature goes down to 50 and 10 K from our current data, which
confirms the connection between the splitting in the optical
conductivity and the FM ordering.

\begin{figure}[t]
\includegraphics[width=2.7in]{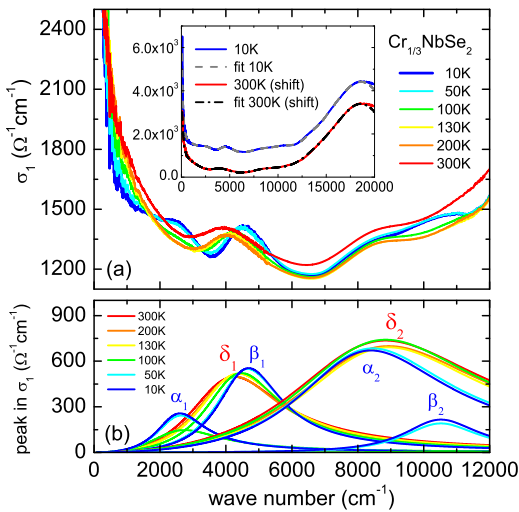}%
\vspace*{-0.20cm}%
\caption{\label{fig:S1}(Color online) (a) $\sigma_1(\omega)$ for
Cr$_{1/3}$NbSe$_2$ below 12000 cm$^{-1}$. The inset illustrates
the experimental data and the Drude-Lorentz fitting for 10 and 300 K
below 20000 cm$^{-1}$. The raw data and the fitting for 300 K are
shifted down by 1000 $\Omega^{-1}$cm$^{-1}$ for clarity. (b) The
first four lorentz terms extracted from the fitting.}
\end{figure}

\section{Theoretical results: the Ground State Band Structure}

Conventionally, the band structure modification after
intercalation is understood by the rigid-band model: the host band
structure is unchanged upon intercalation, meanwhile, electrons
from the guest species fill the host conduction band, thus raise
the chemical potential. However, our optical data apparently
violate the rigid-band model for the emergence of new interband
transitions in Cr doped NbSe$_2$ and their further splitting at
low \emph{T}. To understand these low energy interband
transitions, details about the band structure for both the pure
and the Cr intercalated compounds are required.

We obtain the ground state electronic structure for NbSe$_2$ and
Cr$_{1/3}$NbSe$_2$ by first-principles calculations. The projected
density of states (PDOS) for 2$H$-NbSe$_2$ is shown in Fig. 5 (a)
and (b). The crystal structure of 2$H$-NbSe$_2$ is the same with
that of Cr$_{1/3}$NbSe$_2$ if Cr atoms were removed. Each Nb is in
trigonal prismatic coordination with six Se neighbors, so the Nb
4\emph{d} states are split into: a$_{1g}$ (\emph{d}$_{z^2}$)
state, e'$_g$ doublet (\emph{d}$_{xy}$; \emph{d}$_{x^2-y^2}$), and
e$_g$ doublet (\emph{d}$_{yz}$; \emph{d}$_{zx}$). Far below the
Fermi energy, the total density of states are of Se 4\emph{p}
character. The tail of this Se band hybridizes with Nb (a$_{1g}$,
e'$_g$) complex and extends above E$_F$. A high density of states
at the Fermi energy consists with the optical observation of a
well-defined plasma edge in 2$H$-NbSe$_2$.

We note that the partially filled Nb conduction band is frequently
referred to as "\emph{d}$_{z^2}$" band in the early studies.
However, recent calculation on 2$H$-NbSe$_2$\cite{NbSe2 band} show
that the half-filled conduction band has \emph{d}$_{z^2}$ symmetry
around the $\Gamma$ point of the Brillouin zone, and
(\emph{d}$_{x^2-y^2}$, \emph{d}$_{xy}$) symmetry around the $K$
point. Similar result is reproduced in Fig. 5 (a), that both
a$_{1g}$ (\emph{d}$_{z^2}$) and e'$_g$ (\emph{d}$_{x^2-y^2}$,
d$_{xy}$) contribute to the PDOS of the half-filled conduction
band.

Figure 5 (c) to (e) shows the PDOS for Cr$_{1/3}$NbSe$_2$ in the
FM ground state. Since Cr is in a trigonally distorted CrSe$_6$
octahedron,\cite{Parkin} the Cr 3\emph{d} bands are split into
t$_{2g}$ (which is further split into a$_{1g}$ and e'$_g$) and
e$_g$ states. In the spin up channel, the Cr e'$_g$ state is
localized and fully occupied; while the Cr a$_{1g}$ state and
e$_g$ doublet have weak itinerant character, and strongly
hybridize with Nb conduction band in the range -0.5$\sim$0.5 eV
across E$_F$. All the spin down Cr 3\emph{d} bands are unoccupied.
From Fig. 5 (g) and (h), one will find that Cr$_{1/3}$NbSe$_2$ is
metallic with a Fermi level crossing in both spin directions.

Comparing with 2$H$-NbSe$_2$, the half-filled Nb conduction band
is now almost filled up in Cr$_{1/3}$NbSe$_2$, with a large
removal of DOS at the Fermi energy, leading to a marked reduction
in the free carrier spectral weight. Although there is a strong
hybridization between Cr t$_{2g}$ ($\uparrow$) and e'$_g$
($\uparrow$) states near E$_F$, the occupied Nb states in the spin
down projection is almost unaffected after intercalation, besides
a shift of the chemical potential.

\begin{figure*}
\includegraphics[width=7in]{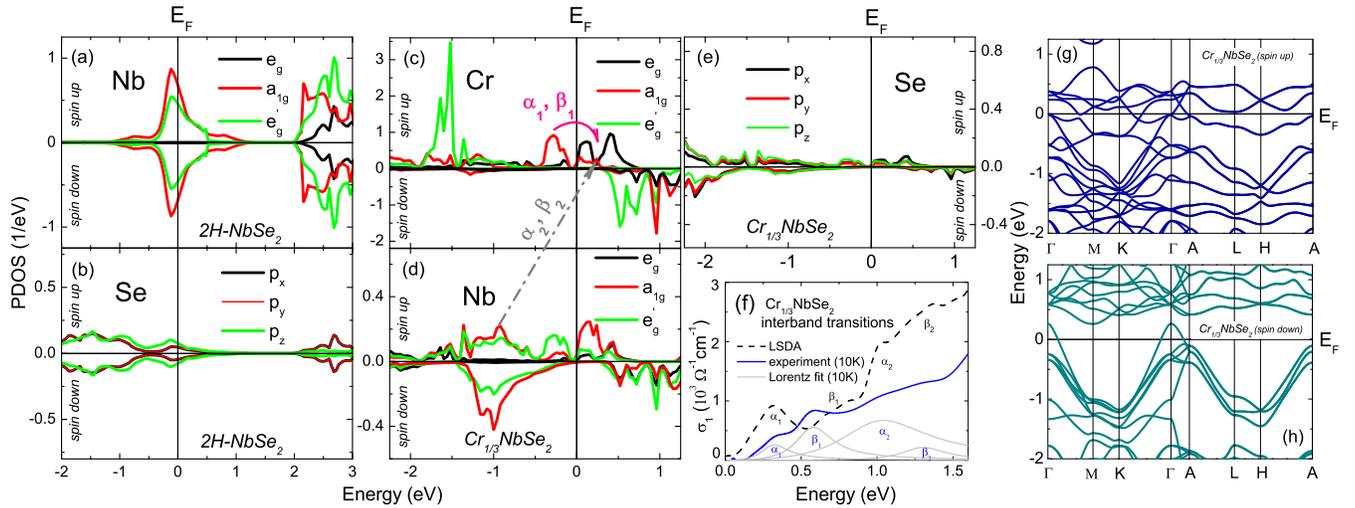}%
\vspace*{-0.20cm}%
\caption{\label{fig:DOS}(Color online) The PDOS for 2$H$-NbSe$_2$:
(a) Nb, and (b) Se. The PDOS for ferromagnetic Cr$_{1/3}$NbSe$_2$:
(c) Cr, (d) Nb,\cite{Nb1} and (e) Se. (f) Calculated interband
transitions for Cr$_{1/3}$NbSe$_2$ in the FM ground state (black
dash line), comparing with the experimental data at 10 K (the
Drude term is subtracted) (blue solid line), and the Lorentz fit
in Fig. 4 (b) (light grey solid line). The band structure for
ferromagnetic Cr$_{1/3}$NbSe$_2$: (g) the spin up, and (h) the
spin down band.}
\end{figure*}

\section{Discussions}

\subsection{The Origin of the Low Energy Interband Transitions}
In comparison with the band structure calculations, we can
identify the origin of the experimental interband transitions
(associated with the joint density of states) at low \emph{T}.

For NbSe$_2$, the close-to-E$_F$ interband transition from
occupied Se to unoccupied Nb states is the origin of the first
peak around 12000 cm$^{-1}$ in Fig. 3 (a). While the second peak
at 20000 cm$^{-1}$ is the Nb \emph{d} to \emph{d} hopping from
E$_F$ to the lowest empty state. Such \emph{d}-\emph{d} transition
is not forbidden with the help of Se 4\emph{p} hybridization.

For Cr$_{1/3}$NbSe$_2$, all possible low-energy ($<$1.5 eV)
interband transitions in the FM ground state include Cr-to-Cr,
Nb-to-Nb, and Cr-to-Nb (also Nb-to-Cr) transitions. Note the PDOS
scales for Fig. 5 (c)-(e) are different that Cr has apparently
larger density of states than that of Nb or Se, so Cr-Cr and Cr-Nb
transitions should dominate the low-energy interband optical
peaks. As shown in Fig. 5 (g), Cr e$_g$($\uparrow$) bands have
rather special dispersion: they are flat near the band bottom and
top but dispersive elsewhere (which can be better resolved along
A-L-H-A), so it has a double-peak PDOS. Then the two peaks,
$\alpha_1$ and $\beta_1$, in $\sigma_1(\omega)$ mainly come from
the spin up Cr a$_{1g}$$\rightarrow$Cr e$_g$ transition, and the
$\alpha_2$ (1.04 eV) and $\beta_2$ (1.30 eV) peaks are mainly
contributed by the occupied spin up Nb (a$_{1g}$,
e'$_g$)$\rightarrow$Cr e$_g$ transition. Other Cr-Cr or Cr-Nb
transitions exceed the energy scope in Fig. 4 (b), while the low
energy Se-Nb transition in NbSe$_2$ shifts to higher energies
because Cr intercalation raises the chemical potential.

The predicted interband contribution in $\sigma_1(\omega)$ from
the LSDA calculation are shown in Fig. 5 (f). The experimental
data with the removal of the Drude component are also presented
for comparison. In general, the theoretical result qualitatively
reproduces the experimental observation of a four-peak character
in the low energy optical response.

\subsection{The Splitting of the Optical Peaks}

Since the Curie-Weiss PM state is a spin-disorder system, that the
orientation and amplitude of neighboring spins would fluctuate
with temperature, one cannot calculate the PM band from
first-principles. However, the band modification across the FM
transition can be speculated from the changes in optical interband
transitions.

As Cr e$_g$($\uparrow$) is the common final state for the four
interband transitions in the FM ground state, then the two-to-four
splitting of the optical peaks should be caused by a splitting, or
more generally, a band broadening, in this common e$_g$ state.
Provided Cr e$_g$($\uparrow$) turns narrower in the PM state, then
the interband transitions $\alpha_i$ and $\beta_i$ are
indistinguishable. The band width for a correlated electronic
system is related to the hopping integral between neighboring
sites, then the long-range ordering of Cr local moment favors a
broader bandwidth from PM to FM state. The spin up Cr a$_{1g}$ and
e$_g$ have partial itinerant character with a PDOS "tail" at E$_F$
(Fig. 5 (c)). Therefore the lowering (i.e. gain) of the kinetic
energy of carriers in the FM transition would lead to the
conduction band broadening.\cite{Hirsch} Here the phase transition
does not apparently change the Drude response in
Cr$_{1/3}$NbSe$_2$, indicating the conducting carriers are mainly
of Nb 4\emph{d} character, thus are less affected by possible
effective mass reduction when the Cr-derived bands turn broadener
in the FM phase.

As shown above, the interband peak splitting is a band broadening
effect in Cr e$_g$ doublet, now further discussions on other
possibilities are necessary.

One possible cause for the band modification is a structure
distortion, which splits the degenerated e$_g$ into two separated
states. Here we should emphasize that the relaxed Se position in
our calculation indicates a trigonal distorted CrSe$_6$ octahedra
(along the c-axis) in the FM ground state, but it apparently does
not change the degeneracy of the e$_g$ doublet (d$_{zx}$;
d$_{yz}$). In fact, the double-peak PDOS for Cr e$_g$($\uparrow$)
is not two separated states, but due to special band dispersion,
so the splitting of the optical peaks is not caused by possible
structural distortion.

Another possible case is the exchange splitting. Regardless of the
detailed band structure for this specific material, the splitting
in the interband transition for a ferromagnet is generally
attributed to the exchange splitting: when magnetism is on the
itinerant side, then the PM phase will have some set of absorption
peaks, and in the FM phase each of them will split into two,
reflecting the fact that the exchange splitting is not rigid.
However, our band structure calculations rule out such a
possibility. As shown in Fig. 5 (c), although the spin up Cr
a$_{1g}$ and e$_g$ have a weak itinerant character, the Cr
3\emph{d} bands are mainly localized, therefore the dominating
magnetic mechanism for Cr$_{1/3}$NbSe$_2$ is away from the
itinerant side. In the Curie-Weiss PM state, the Cr 3\emph{d}
local moments do not vanish. Then the only close-to-E$_F$ band
which might experience spin polarization from the PM to FM state
is the Nb 4\emph{d} conduction band (Fig. 5 (d)). However, one can
not find any Nb-related interband transitions which satisfy: 1) a
0.5$\sim$0.7 eV and 1$\sim$1.2 eV gap for both spin projections;
and 2) a 0.25 eV energy difference for the common hopping in
different spin channels (as a 2000 cm$^{-1}$ splitting from
optical measurements). In fact, the magnetic moment for Nb atom is
less than 0.03$\mu_B$ (while Cr moment is 2.6$\mu_B$) from our
calculation, hence the Nb spin polarization is so weak that it can
be neglected, so the splitting of the interband transitions is
irrelevant to the exchange splitting.

Here some complementary discussion on the PM band structure should
also be added. In the Curie-Weiss paramagnetic state, there is no
spin polarization for the energy band, then the optical response
from different spin channels have equal contribution to
$\sigma_1(\omega)$. As shown above, we conclude that the "spin up"
Cr e$_g$ turns narrower in the PM state. Here the spin label was
used to better describe this specific band when comparing with the
FM case. One should keep in mind that the two spin channels are
mixed in the PM state, reflecting the fluctuation of the local
spins. However, such a band mixing will not bring new low-energy
interband transitions. Note $\alpha_i$ and $\beta_i$ (i=1,2) are
all come from Cr-related bands, which are mainly localized and has
a large splitting in different spin projections in the FM ground
state (Fig. 5 (c)); furthermore, possible spin polarization in the
FM state for the Nb and Se related bands are rather weak, then the
band mixing in the PM state will not bring new peaks for the low
frequency $\sigma_1(\omega)$.

Finally, it should be noted that the qualitative electronic
structure around Fermi level for Cr$_{1/3}$NbSe$_2$ is not
sensitive to the on-site Coulomb repulsion $U$ of Cr site due to
the strong Cr-Nb hybridization. We have tried a LSDA+$U$ ($U$=3.0
eV for Cr site) calculation, the low energy optical conductivity
is not qualitatively changed (except the $\alpha_1$ and $\beta_1$
peaks are weaker compared to LSDA).

\section{Conclusion}
We report a combined optical spectroscopy study and
first-principles calculations on 2$H$-type NbSe$_2$ and the
ferromagnetic intercalated compound Cr$_{1/3}$NbSe$_2$. A weak
mid-infrared suppression in R($\omega$) of 2$H$-NbSe$_2$ testifies
the existence of a CDW induced partial gap on the Fermi surface.
For Cr$_{1/3}$NbSe$_2$, the appearance of new interband
transitions and their remarkable \emph{T} evolution are direct
experimental evidence for the invalidity of the rigid-band model.
The multi-peak feature in $\sigma_1$($\omega$) has been clearly
extracted by a Drude-Lorentz fitting. Based on the LSDA
calculation, we found that the four optical peaks in the FM phase
are interband transitions involving both the intercalated Cr
3\emph{d} and the host Nb 4\emph{d} states. Considering the
temperature evolution for these optical interband transitions, we
further conclude that their splitting from PM to FM phase is not
an exchange splitting effect, but a dispersion modification
(broadening) in the spin up Cr $e_g$ doublet, which have partial
itinerant character and strongly hybridize with Nb 4\emph{d}
states. Such an evident band broadening in a rather close-to-E$_F$
3\emph{d} state requires further experimental investigations,
which might provide new insight into understanding the
ferromagnetism induced band modification across the phase
transition.

\begin{acknowledgments}
This work is supported by the National Science Foundation of
China, the Knowledge Innovation Project of the Chinese Academy of
Sciences, and the 973 project of the Ministry of Science and
Technology of China. The work at Brookhaven National Laboratory is
operated for the U.S. Department of Energy by Brookhaven Science
Associates (No. DE-Ac02-98CH10886), and at Princeton by the
National Science Foundation of the USA.
\end{acknowledgments}

\end{document}